\documentclass[showpacs,prl,twocolumn,floatfix,
nofootinbib,superscriptaddress]{revtex4}
\usepackage{bm}
\usepackage{epsf}

\newcommand{\beq}{\begin{equation}}
\newcommand{\eeq}{\end{equation}}
\newcommand{\bea}{\begin{eqnarray}}
\newcommand{\eea}{\end{eqnarray}}

\newcommand{\mc}[3]{\multicolumn{#1}{#2}{#3}}
\newcommand{\order}[1]{O(#1)}  
\newcommand{\wtilde}[1]{\widetilde{#1}}
\newcommand{\pslash}[1]{\rlap{/}\kern-0.8pt #1}
\newcommand{\Dslash}{\rlap{/}\kern-2.0pt D}

\newcommand{\LambdaQCD}{\Lambda_{\rm QCD}}
\def\today{\number\day\space\ifcase\month\or
January\or February\or March\or April\or May\or June\or
July\or August\or September\or October\or November\or December\fi
\space\number\year}
\newcount\mins \newcount\hours
\def\now{\hours=\time \mins=\time
	\divide\hours by60 \multiply\hours by60 \advance\mins by-\hours
	\divide\hours by60 
	\number\hours:\ifnum\mins<10 0\fi\number\mins}

\def\stacksymbols #1#2#3#4{\def\theguybelow{#2}
    \def\verticalposition{\lower#3pt}
    \def\spacingwithinsymbol{\baselineskip0pt\lineskip#4pt}
    \mathrel{\mathpalette\intermediary#1}}
\def\intermediary #1#2{\verticalposition\vbox{\spacingwithinsymbol
    \everycr={}\tabskip0pt
    \halign{$\mathsurround0pt#1\hfil##\hfil$\crcr#2\crcr
             \theguybelow\crcr}}}

\newcommand{\nl}{\nonumber \\}

\begin{document}


\title{The $\bm{B_s}$ and $\bm{D_s}$ decay constants in 3~flavor lattice QCD}

\author{Matthew Wingate}
\affiliation{Department of Physics,
The Ohio State University, Columbus, OH 43210, USA}
\affiliation{Institute for Nuclear Theory, University of Washington,
Seattle, WA 98195-1550, USA}
\author{Christine T.\ H.\ Davies}
\affiliation{Department of Physics \& Astronomy,
University of Glasgow, Glasgow, G12 8QQ, UK}
\author{Alan Gray}
\affiliation{Department of Physics \& Astronomy,
University of Glasgow, Glasgow, G12 8QQ, UK}
\affiliation{Department of Physics,
The Ohio State University, Columbus, OH 43210, USA}
\author{G.\ Peter Lepage}
\affiliation{Laboratory of Elementary Particle Physics,
Cornell University, Ithaca, NY 14853, USA}
\author{Junko Shigemitsu}
\affiliation{Department of Physics,
The Ohio State University, Columbus, OH 43210, USA}

\date{\today}


\begin{abstract}
Capitalizing on recent advances in lattice QCD, we present a calculation
of the leptonic decay constants $f_{B_s}$ and $f_{D_s}$ that includes
effects of one strange sea quark and two light sea quarks.  The
discretization errors of improved staggered fermion actions are small 
enough to simulate with 3 dynamical flavors on lattices with spacings
around 0.1 fm using present computer resources.  By shedding the
quenched approximation and the associated lattice scale ambiguity,
lattice QCD greatly increases its predictive power.  NRQCD is used to
simulate heavy quarks with masses between $1.5m_c$ and $m_b$.  
We arrive at the following results: 
$f_{B_s} = 260 \pm 7 \pm 26 \pm 8 \pm 5$ MeV and 
$f_{D_s} = 290 \pm 20 \pm 29 \pm 29 \pm 6$ MeV.  
The first quoted error is the statistical uncertainty, and the 
rest estimate the sizes of higher order terms neglected in 
this calculation.  All of these uncertainties are systematically improvable
by including another order in the weak coupling expansion, the nonrelativistic
expansion, or the Symanzik improvement program.
\end{abstract}

\pacs{12.38.Gc,
13.20.Fc, 
13.20.He } 

\maketitle


We present the first complete calculation of the $B_s$ and $D_s$ decay
constants with 3 flavors of sea quarks with small masses.  
The $D_s$ decay constant, $f_{D_s}$, has been measured 
\cite{Chadha:1998zh,Heister:2002fp}, and CLEO-c promises to reduce
the experimental errors significantly.  The comparison of
experimental and lattice results will be a vital test of
lattice QCD.  The $B_s$ and $B_d$ decay constants, $f_{B_s}$ and $f_B$,
are necessary in order to constrain $V_{td}$ via $\overline{B^0}-B^0$ mixing.  
Since experimental measurements of $f_{B_s}$ remain elusive,
an accurate lattice calculation is critical 
for improving phenomenological tests of CKM unitarity.

The calculation presented below makes use of lattice Monte Carlo simulations,
done by the MILC collaboration \cite{Bernard:2001av},
which include the proper sea quark content: one dynamical strange quark
and two flavors of dynamical quarks with masses as light as $m_s/4$.
The correct number of flavors is necessary in order for the
lattice theory to have the same $\beta$-function as real QCD.
Only then can we expect, in principle, lattice results to agree with
experimental measurements.

Inclusion of light up and down sea quarks
is essential for accurate lattice phenomenology.  
The innovation which allows this on present computers is an improved 
staggered discretization of the light quark action
\cite{Bernard:1998gm,Lepage:1998id,Bernard:1998mz,
Lepage:1998vj,Orginos:1998ue,Toussaint:1998sa,Orginos:1999cr,Bernard:1999xx}.
For each dynamical flavor the fourth root of the fermion
determinant is used in the Monte Carlo updating algorithm.  
Despite some open theoretical issues concerning staggered fermion algorithms, 
these calculations are free of ambiguities
present in quenched simulations, provided the
$u$ and $d$ sea quark masses are light enough.
In practice, calculations involving
unstable particles or multi-hadronic final states require new or
refined techniques before they can be reliably simulated; however, there
are several quantities for which numerical calculation is straightforward.
Ref.~\cite{Davies:2003ik} presents lattice results for a variety of 
these ``gold-plated'' quantities,  including $f_\pi$ and $f_K$, and
several splittings in the $\Upsilon$ spectrum.
The change from $n_f=0$ results, which differ substantially from 
experiment, to $n_f=3$ results, which agree with experiment up to
the 3\% lattice uncertainties, suggests that the improved staggered
fermion method correctly simulates QCD.
Heavy-light pseudoscalar leptonic decay constants, presented here,
fit into the ``gold-plated'' category \cite{Davies:2003ik}.

Our calculation of $f_{B_s}$ and $f_{D_s}$ uses standard lattice QCD methods.
Correlation functions were computed using a subset of gauge field
configurations generated by the MILC collaboration.
The configurations include
the effects of 2 dynamical light quarks with equal bare mass, denoted by
$m_\ell^\mathrm{sea}$,
and 1 dynamical strange quark, with bare mass $m_s^\mathrm{sea}$.
The lattices we used have spacings of
about 1/8 fm and volumes of about $(2.5\;\mathrm{fm})^3\times 8.0$ fm.
We focus on the configurations where
$m_\ell^\mathrm{sea}/m_s^\mathrm{sea} = 1/5$ and 2/5.
As we discuss below, the physical strange quark mass, $m_s$, obtained
from the light hadron spectrum is actually $4/5$ of $m_s^\mathrm{sea}$, 
so the light sea quark masses are approximately $m_s/4$ and $m_s/2$.  

Details of the Monte Carlo simulations which generated the
ensemble of gauge fields are given in \cite{Bernard:2001av}. 
In this work we use a level splitting in the $\Upsilon$ spectrum,
e.g.\ the $\Upsilon(\mathrm{2S-1S})$ splitting, to determine
the lattice spacing, instead of the length scale $r_1$, derived from
the static quark potential, which was used in \cite{Bernard:2001av}. 
The $\Upsilon$ spectrum will be presented in detail in a future 
publication \cite{Gray:UpsSpect}; 
however, we note here that as the light sea quark mass
is increased, the $\Upsilon(\mathrm{1P-1S})$ splitting becomes 
slightly smaller than experiment when the lattice spacing is set by
$\Upsilon(\mathrm{2S-1S})$.  With statistical errors between $1-2\%$,
the difference between experiment and lattice values for the 
$\mathrm{1P-1S}$ splitting is $0\sigma$, $1\sigma$, and $2\sigma$ for 
$m_\ell^\mathrm{sea}/m_s \approx$ 1/4, 1/2, and 3/4, respectively.
Except for estimating sea quark mass effects, we use the 
$m_\ell^\mathrm{sea}/m_s \approx 1/4$ lattice to obtain our results.


\begin{table}
\caption{\label{tab:heavymass}Heavy quark parameters and meson masses
for two values of the light sea quark mass.}
\begin{center}
\begin{tabular}{cc|ll|ll}
&&\mc{2}{c}{$m_\ell^\mathrm{sea}\approx m_s/4$ (568 config)} & 
\mc{2}{c}{$m_\ell^\mathrm{sea}\approx m_s/2$ (468 config)} \\
$aM_0$ & $n$ & \mc{1}{c}{$m_{H_s}^\mathrm{kin}$ (GeV)} &
\mc{1}{c}{$m_{H_s}^\mathrm{pert}$ (GeV)} &
\mc{1}{c}{$m_{H_s}^\mathrm{kin}$ (GeV)} &
\mc{1}{c}{$m_{H_s}^\mathrm{pert}$ (GeV)} \\ \hline
2.8 & 2 & ~~5.6(2)   & ~~5.3(4)   & ~~5.22(17)  & ~~5.4(4) \\
2.1 & 4 & ~~4.38(10) & ~~4.2(3)   & ~~4.28(11)  & ~~4.3(3) \\
1.6 & 4 & ~~3.52(6)  & ~~3.5(3)   & ~~3.56(7)   & ~~3.5(3) \\
1.2 & 6 & ~~2.84(5)  & ~~2.78(19) & ~~2.93(4)   & ~~2.83(19) \\
1.0 & 6 & ~~2.53(4)  & ~~2.41(16) & ~~2.60(3)   & ~~2.45(16)
\end{tabular}
\end{center}
\end{table}

The analysis of \cite{Bernard:2001av} has also been updated
with respect to determining the quark mass corresponding to the
physical strange quark mass sector.  Rather than using $\bar{s}s$
mesons which are either unstable (the $\phi$) or do not exist
(the pseudoscalar) we use the results of the partially quenched chiral
perturbation theory analysis of $m_K$ and $m_\pi$ 
\cite{Davies:2003ik,Aubin:2002ss}. 
On lattices with $au_0 m_\ell^\mathrm{sea}=0.007, 0.01$, and $0.02$,
the physical strange sector is obtained with
$a u_0 m_s^\mathrm{val} = 0.040$.
(Here we keep explicit a factor of the
mean-link $u_0$, which was absorbed into the definition
of the light bare quark mass in \cite{Bernard:2001av}.)

Table~\ref{tab:heavymass} lists the bare heavy quark masses
used in this work, along with the corresponding NRQCD stabilization 
parameter $n$.  For each mass, we estimate
the heavy-light meson mass two ways: the ``kinetic mass'' 
$m_{H_s}^\mathrm{kin}$ is extracted from finite momentum correlators
using the meson dispersion relation, and the ``perturbative mass''
$m_{H_s}^\mathrm{pert}$ is estimated from the zero momentum correlator 
and the one-loop heavy quark mass renormalization.  We find that
the bare heavy quark mass $aM_0 = 2.8$ produces the correct 
experimental mass for the $\Upsilon$ \cite{Gray:2002vk} and the $B_s$
within statistical errors.

When we construct correlation functions for heavy-light
mesons, we use the equivalence between staggered and naive
fermion propagators.  The operators we use are the
same ones as with Wilson-like discretizations or in the continuum limit.
This method for computing the $B_s$ mass
and decay constant has been presented, along with tests on
quenched lattices, in recent work \cite{Wingate:2002fh}.

The $B_s$ decay constant is defined through the axial vector matrix
element; 
$\langle 0 | A_\mu | B_s(p_\mu) \rangle ~=~ f_{B_s} p_\mu$.
Here we use only the temporal component.  
Up through $\order{1/M_0}$ three lattice operators contribute 
to $A_0$, the leading-order current, $J_0$, and two sub-leading currents
$J_1$ and $J_2$;
the matching from the lattice to the continuum 
is done in perturbation theory \cite{Morningstar:1998ep}.
A one-loop calculation yields
\beq
{A}_0 = (1+\alpha_s\wtilde{\rho}_0) J_0^{(0)}\! +\!
(1 + \alpha_s\rho_1) J_0^{(1,\mathrm{sub})}
\! + \alpha_s\rho_2 J_0^{(2,\mathrm{sub})} 
\label{eq:A0final2}
\eeq
where $J_0^{(1,\mathrm{sub})}= J_0^{(1)}
~-~ {\alpha_s \zeta_{10}} J_0^{(0)}$,
and so too for $J_0^{(2,\mathrm{sub})}$.  These operators
explicitly subtract, through one-loop, the power-law mixing
of $J_0^{(0)}$ with $J_0^{(1)}$ and  $J_0^{(2)}$
\cite{Collins:2000ix}.
The perturbative calculation which determines $\wtilde{\rho}_0,
\rho_1, \rho_2, \zeta_{10}$, and $\zeta_{20}$
will be presented separately \cite{Gulez:HLPT}.




\begin{figure}
\begin{center}
\epsfxsize=\hsize
\vspace{-1cm}
\mbox{\epsfbox{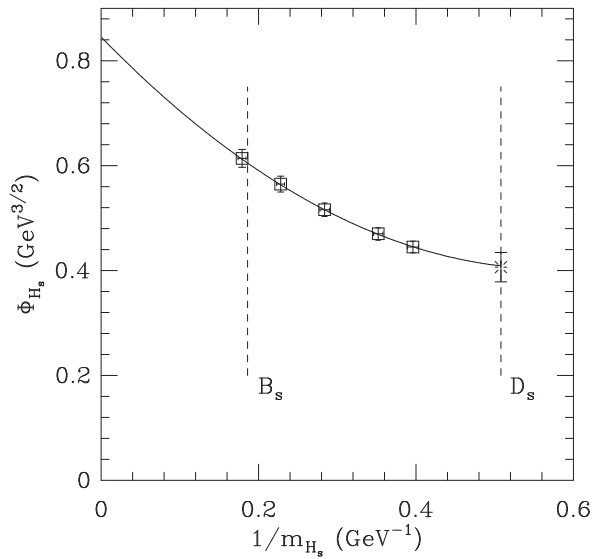}}
\caption{Squares show $\Phi_{H_s} \equiv f_{H_s}\sqrt{m_{H_s}}$ 
vs.\ $1/m_{H_s}$ on the $m_\ell^\mathrm{sea}/m_s \approx 1/4$ lattice.
The solid line shows the fit described in the text, and the asterisk
shows the value of $\Phi_{H_s}$ extrapolated to the charm sector.
Experimental values for $1/m_{B_s}$ and $1/m_{D_s}$ are shown
as dashed vertical lines.}
\label{fig:PhiHs_vs_invhl}
\end{center}
\end{figure}

From fits to correlation functions, we extract the combination
$\Phi_{H_s} \equiv f_{H_s}\sqrt{m_{H_s}}$.  Let us denote by $\Phi^{(i)}$ 
the contribution of $J_0^{(i)}$ to $\Phi_{H_s}$. 
For this calculation the matrix
elements are computed for mesons at rest, so $\Phi^{(1)}= \Phi^{(2)}$.
Table~\ref{tab:Phi} summarizes fits to the numerical data, converted
to physical units using $1/a = 1.59(2)$ GeV for the
$m_\ell^\mathrm{sea} \approx m_s/4$ lattice and
$1/a=1.61(2)$ GeV for the $m_\ell^\mathrm{sea} \approx m_s/2$ lattice.
(The quoted uncertainties come from statistical and fitting uncertainties
in the $\Upsilon(\mathrm{2S-1S})$ splitting \cite{Gray:UpsSpect}.)
By comparing $\Phi^{(1)}/\Phi^{(0)}$ with $\Phi^{(1,\mathrm{sub})}/\Phi^{(0)}$
one can observe the sizable power law mixing of $J^{(0)}$ with $J^{(1)}$.
The expression (\ref{eq:A0final2}) absorbs the mixing back into the
term proportional to $J^{(0)}$, so $\Phi^{(1,\mathrm{sub})}/\Phi^{(0)}$
represents the physical contribution of $1/M_0$ terms to
$\Phi_{H_s}$ up to two-loop corrections.  
The 4\% contribution from the $1/M_0$ operator we see for the $B_s$ 
($aM_0=2.8$) is the same size seen in quenched studies over a range of 
lattice spacings \cite{Collins:2000ix}.


\begin{table}[t]
\caption{\label{tab:Phi}Simulation results for $\Phi_{H_s}\equiv 
f_{H_s}\sqrt{m_{H_s}}$ for each light sea quark mass and
heavy quark mass.  The third column lists the leading order
term $\Phi^{LO}\equiv (1+\alpha_s\wtilde{\rho}_0)\Phi^{(0)}$,
The fourth and fifth columns show 
the contributions of $J_0^{(1)}$, with respect to $\Phi^{(0)}$,
before and after the power law subtraction.  
The sixth column gives the result for $\Phi_{H_s}$.
Statistical and fitting uncertainties are quoted in parentheses
(not including the statistical uncertainty in $1/a$).}
\begin{center}
\begin{tabular}{ccccc}
\mc{5}{c}{$m_\ell^\mathrm{sea}/m_s\approx 1/4$} \\ 
$aM_0$ & $\Phi^{LO}$(GeV${}^{3/2}$) & $\Phi^{(1)}/\Phi^{(0)}$ & 
$\Phi^{(1,\mathrm{sub})}/\Phi^{(0)}$  & 
$\Phi_{H_s}$(GeV${}^{3/2}$) \\ \hline
      2.8 & 0.640(11) & ~~$-9.0(4)\,\%$ & $-3.7(4)\,\%$ & 0.614(13)\\
      2.1 & 0.598(10) & $-11.7(4) \,\%$ & $-5.0(4)\,\%$ & 0.565(11)\\
      1.6 & 0.557(8)  & $-14.7(4) \,\%$ & $-6.4(4)\,\%$ & 0.516(9) \\
      1.2 & 0.519(7)  & $-18.3(4) \,\%$ & $-7.8(4)\,\%$ & 0.470(8) \\
      1.0 & 0.499(6)  & $-20.7(4) \,\%$ & $-8.6(4)\,\%$ & 0.445(7) \\ \hline
\mc{5}{c}{$m_\ell^\mathrm{sea}/m_s\approx 1/2$} \\
$aM_0$ & $\Phi^{LO}$(GeV${}^{3/2}$) & $\Phi^{(1)}/\Phi^{(0)}$ & 
$\Phi^{(1,\mathrm{sub})}/\Phi^{(0)}$  & 
$\Phi_{H_s}$(GeV${}^{3/2}$) \\ \hline
      2.8 & 0.640(15) & ~~$-8.9(6)\,\%$ & $-3.6(6)\,\%$ & 0.615(14) \\
      2.1 & 0.599(11) & $-11.4(6) \,\%$ & $-4.7(6)\,\%$ & 0.567(12) \\
      1.6 & 0.563(7)  & $-14.2(5) \,\%$ & $-5.9(5)\,\%$ & 0.523(9) \\
      1.2 & 0.528(6)  & $-17.7(5) \,\%$ & $-7.1(5)\,\%$ & 0.481(6) \\
      1.0 & 0.510(6)  & $-20.0(5) \,\%$ & $-7.9(5)\,\%$ & 0.459(6)
\end{tabular}
\end{center}
\end{table}

Figure~\ref{fig:PhiHs_vs_invhl} shows the heavy quark mass dependence
of $\Phi_{H_s}$ on $m_{H_s}$ (plotted as squares).  The data are fit well by 
\beq
\Phi_{H_s} ~=~ \Phi_{H_s}^\mathrm{stat} \left( 1 + \frac{C_1}{m_{H_s}} 
+ \frac{C_2}{m_{H_s}^2}\right)
\label{eq:PhiVsHQ}
\eeq
with a correlated $\chi^2$ per degree-of-freedom of $0.7/2$.
We find $\Phi_{H_s}^\mathrm{stat} = 0.85(4)$ GeV${}^{3/2}$,
$C_1 = -1.82(20)$ GeV, and $C_2 = 1.59(35)~\mathrm{GeV}^2$.
It is important to note that most of the mass dependence of $\Phi_{H_s}$
comes through the action and not through the $1/M_0$ currents.
A fit to the leading order 
$\Phi^{LO}\equiv (1+\alpha_s\wtilde{\rho}_0)\Phi^{(0)}$ 
(Table~\ref{tab:Phi}) yields similar fit parameters:
$\Phi^{\mathrm{stat},LO} = 0.83(4)$ GeV${}^{3/2}$,
$C_1^{LO} = -1.53(19)$ GeV, and $C_2^{LO} = 1.32(32)~\mathrm{GeV}^2$.
Since the action is responsible for most of the $m_{H_s}$ dependence
and is accurate through $\order{\LambdaQCD^2/m_Q^2}$,
performing the operator matching
through $\order{1/M_0^2}$ should yield only small corrections
to the values for $\Phi_{H_s}^\mathrm{stat}$, $C_1$, and $C_2$ listed above. 

We use the fit above to interpolate $\Phi_{H_s}$ slightly to the physical 
value of $1/m_{B_s}$.  The result is $f_{B_s} = 260(7)$ MeV, where
the quoted statistical error combines statistical errors from 
$1/a$ and $a^{3/2}\Phi_{B_s}$.
Uncertainties in $m_{H_s}^\mathrm{kin}$ are small compared to the 
other uncertainties in the fit.
The systematic uncertainties due to the neglect of higher order terms
are estimated by assuming coefficients of $\order{1}$.  For example,
two-loop terms omitted in (\ref{eq:A0final2}) are estimated to be
10\% effects, taking the coupling constant defined through the 
plaquette $\alpha_s = \alpha_P^{n_f=3}(2/a) = 0.32$ \cite{Davies:2002mv}.
This is the largest systematic uncertainty.  Leading discretization
errors are $\order{\alpha_s a^2 \LambdaQCD^2}$, where $\LambdaQCD$ is
the typical scale of nonperturbative dynamics.  Taking 
$1/a = 1.6$ GeV and $\LambdaQCD = 400$ MeV implies 2\% cutoff effects.
The operator matching neglects terms $\order{\LambdaQCD^2/m_b^2}$ or
approximately 1\%.  We note that
the coefficient of the spin-dependent term in the action is the tree level
one, giving rise to an $\order{\alpha_s\LambdaQCD/m_Q}$ error in the action,
which is about 3\% for bottom quarks.  This does not necessarily
translate into a comparably large error in decay constants where the 
spin dependence is not obvious; however, to be conservative, we 
quote an overall 3\% uncertainty due to relativistic corrections.

Extrapolating the fit described above to the physical value for $1/m_{D_s}$ 
requires care since higher order terms in $1/m_{H_s}$ become increasingly
important.  The fit above extrapolates to $f_{D_s} = 290(10)$ MeV.  We
use a Bayesian analysis to estimate possible effects due to higher
order terms.  Allowing terms like $C_n/m_{H_s}^n$ with $n\ge 3$ in the
fit (\ref{eq:PhiVsHQ}), with Gaussian priors for $C_n/\mathrm{GeV}^n
= 0 \pm \delta$ and $\delta = 1-4$, 
we find higher order terms could lead to a 20 MeV
error in $f_{D_s}$.   This is taken as the statistical and fitting 
uncertainty.  The $\order{\alpha_s^2}$ perturbative error
and the $\order{\alpha_s a^2 \LambdaQCD^2}$ discretization error
are again estimated to be 10\% and 2\%, respectively. 
For charmed mesons $\LambdaQCD/m_c = 1/4$, so
the $\order{\LambdaQCD^2/m_Q^2}$
corrections to the matching are estimated to be 6\%.  As mentioned above,
the $\order{\alpha_s\LambdaQCD/m_Q}$ error in the spin dependent term in
the action probably does not lead to a proportional error in the decay 
constant.
Nevertheless, we quote 10\% as a conservative estimate of possible 
relativistic corrections. 


\begin{figure}
\begin{center}
\epsfxsize=\hsize
\vspace{-3cm}
\mbox{\epsfbox{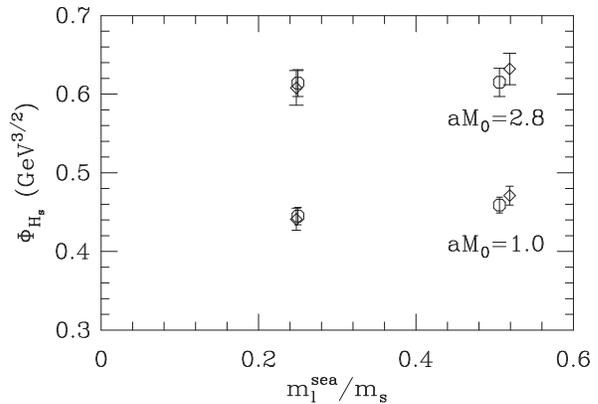}}
\caption{Dependence of $\Phi_{H_s}$ on the light sea quark mass.
The upper data points correspond to the $B_s$ ($aM_0 =2.8$) and
the lower points come from our lightest heavy quark mass, $aM_0 = 1.0$.
The different symbols indicate which quantity was used to set the 
lattice spacing: octagons use the $\Upsilon(\mathrm{2S-1S})$ splitting and
diamonds use the $\Upsilon(\mathrm{1P-1S})$ splitting.
Error bars represent combined statistical uncertainties of 
$1/a$ and $a^{3/2}\Phi_{H_s}$.}
\label{fig:PhiHs_vs_rsea}
\end{center}
\end{figure}

Figure~\ref{fig:PhiHs_vs_rsea} shows $\Phi_{H_s}$ vs.\ 
light sea quark mass for two heavy quark masses.  We show
the results where the lattice
spacing is set from the $\Upsilon(\mathrm{2S-1S})$ splitting
and the same data adjusted as if the spacing were set from the 
$\Upsilon(\mathrm{1P-1S})$ splitting.  No sea quark mass dependence
is observed for $aM_0=2.8$, nor is a significant dependence
observed for $aM_0=1.0$ when the spacing is set with 
$\Upsilon(\mathrm{2S-1S})$.  Only because there is a 2\%
ambiguity in setting the scale with $m_\ell^\mathrm{sea}\approx m_s/2$
does there appear to be some sea quark mass effect in the $aM_0=1.0$
data with the $\Upsilon(\mathrm{1P-1S})$ spacing.  No such 
scale setting ambiguity
exists for the data with $m_\ell^\mathrm{sea}\approx m_s/4$, which
is where our main results are obtained.  

Experimental measurements of the leptonic decay constants for
$B_s$ and $D_s$ are challenging, so these are rare cases where
lattice QCD can lead experiment.
The most recent and precise experimental results for $f_{D_s}$ 
are $280\pm 17\pm 25\pm 34$ MeV \cite{Chadha:1998zh} and 
$285\pm 19\pm 40$ MeV \cite{Heister:2002fp}, which agree well with
the calculation presented here.  Shrinking the experimental uncertainties
on $f_{D_s}$ is a major goal of the CLEO-c program.

Turning to comparison with the existing literature on lattice QCD,
we quote recent world averages and comment on another new result.
The $n_f = 3$ results presented here have central values larger than 
lattice calculations with $n_f = 0$ or 2.  For example, recent
averages \cite{Ryan:2001ej} of quenched results are 
$f_{B_s}^{n_f=0} = 200(20)$ MeV and $f_{D_s}^{n_f=0} = 230(14)$ MeV,
significantly lower than our results.  The 2 flavor world averages,
$f_{B_s}^{n_f=2} = 230(30)$ MeV and $f_{D_s}^{n_f=2} = 250(30)$ MeV,
are higher than the quenched and agree within the quoted 
uncertainties.

Recently the JLQCD collaboration reported a lattice result
using two dynamical flavors of improved Wilson fermions with mass between
$0.7m_s$ and $2.9m_s$ \cite{Aoki:2003xb}.  They quote
$f_{B_s} = 215(9){+0\choose -2}(13){+6\choose -0}$ MeV,
with the first error statistical, second from chiral extrapolation
of the sea quark mass, third from finite lattice spacing
combined with truncation of NRQCD and perturbative expansions,
and the fourth from ambiguity in setting the strange quark mass.
This calculation uses much larger quark masses than ours, uses unstable
hadron masses to set the lattice spacing and $m_s$, and does not include
a dynamical strange quark.  
Further work will be required to determine which, if any,
of these accounts for the differences with our results.
For instance, it
would be interesting to calculate $a^{-1}$ from several $\Upsilon$
splittings on the configurations of \cite{Aoki:2003xb} 
and compare $f_{B_s}$ using those scales to the value quoted 
in their paper, which uses the scale set by $m_\rho$. 

Recent sum rule calculations agree within errors for the $B_s$ 
decay constant: e.g.\ $f_{B_s}^\mathrm{s.r.} = 236(30)$ MeV 
\cite{Jamin:2001fw,Narison:2001pu}.
On the other hand, they do not calculate an increase in the decay constant
as the heavy quark mass decreases:
e.g.\ $f_{D_s}^\mathrm{s.r.} = 235(24)$ MeV \cite{Narison:2001pu}. 

To summarize, we have completed a calculation of the $B_s$ and $D_s$ 
decay constants using 3 flavor lattice QCD.  
The more realistic sea quark content
allows a unique lattice spacing to be determined using one of
several quantities \cite{Davies:2003ik}, enabling more reliable
prediction of quantities not yet measured experimentally.  
Our final results are
\bea
f_{B_s} &~=~& 260 ~\pm~ 7 ~\pm~ 26
	~\pm~ ~8 ~\pm~ 5 ~\mathrm{MeV}\nl
f_{D_s} &~=~& 290 ~\pm~ 20 ~\pm~ 29
	~\pm~ 29 ~\pm~ 6 ~\mathrm{MeV} \, .
\label{eq:finalresult}
\eea
The uncertainties quoted are respectively due to statistics and fitting,
perturbation theory, relativistic corrections, and discretization effects.
The result for the $D_s$ decay constant agrees with experimental 
determinations, and the result for the $B_s$ decay constant is a 
prediction for future experiments.  Improvement of the lattice results
requires a two-loop perturbative matching calculation or the use of
fully nonperturbative methods.  On the other hand, much of the perturbative
uncertainty cancels in the ratio $f_{B_s}/f_{B_d}$.  Work is underway
to study $f_{B_d}$ using the methods discussed in this paper 
\cite{Wingate:2003ni}.

Simulations were performed at NERSC.  We thank the MILC
collaboration for their gauge field configurations.
This work was supported in part by the DOE, NSF, PPARC, and the EU.

\bibliography{mbw}
\bibliographystyle{apsrev}


\end{document}